\documentclass[conference]{IEEEtran}
\IEEEoverridecommandlockouts
\IEEEpubid{\makebox[\columnwidth]{979-8-3503-8481-9/24/\$31.00~\copyright2024 IEEE\hfill} \hspace{\columnsep}\makebox[\columnwidth]{ }}
\usepackage{cite, balance}
\usepackage{amsmath,amssymb,amsfonts}
\usepackage{algpseudocode}
\usepackage{graphicx}
\usepackage{textcomp}
\usepackage{xcolor}
\def\BibTeX{{\rm B\kern-.05em{\sc i\kern-.025em b}\kern-.08em
    T\kern-.1667em\lower.7ex\hbox{E}\kern-.125emX}}
\usepackage[linesnumbered,ruled]{algorithm2e}
    
\begin{document}

\title{The Error Analysis of the Secret Key Generation Algorithm Using Analog Function Computation}

\author{
}

\author{\IEEEauthorblockN{Ertugrul Alper\normalsize $^{1}$, Eray Guven\normalsize $^{2}$, Gunes Karabulut Kurt\normalsize $^{2}$, Enver Ozdemir\normalsize $^{1}$} 
\IEEEauthorblockA{{\normalsize $^{1}$Informatics Institute, Istanbul Technical University, Istanbul, Turkey}\\
{\normalsize $^{2}$Poly-Grames Research Center, Department of Electrical Engineering,  Polytechnique Montr\'eal, Montr\'eal, Canada}
}
E-mail: alper16@itu.edu.tr, guven.eray@polymtl.ca, gunes.kurt@polymtl.ca, ozdemiren@itu.edu.tr
}

\maketitle
\IEEEpubidadjcol
\begin{abstract}
This study introduces a decentralized approach to secure wireless communication using a cryptographic secret key generation algorithm among distributed nodes. The system model employs Gaussian prime numbers, ensuring the collaborative generation of a secret key. Pre-processing and post-processing functions enable to generate a secret key across the network. An error model evaluates aspects like thermal noise power and channel estimation errors, while simulations assess the success rate to factorize the norm of the secret key. It is observed that path loss-induced large scale fading emerges as a critical component impacting information and power loss. The robustness of the proposed model under fading channel conditions is evaluated with a success rate. Additionally, it is also observed that the tolerance value set in the factorization algorithms has a significant impact on the success rate. Furthermore, the success rate is compared in two scenarios, one with 2 users and another with 3 users, to provide a comprehensive evaluation of the system performance. 
\end{abstract}


\section{Introduction}

In the dynamic world of wireless communication, it is crucial to ensure and enhance information exchange between connected nodes. In this study, a comprehensive system model is presented, encompassing a distributed network of $N$ nodes, which operates in a full-duplex communication environment over the air.  Notably, the ever-present challenge of thermal noise power in the medium is contended.

In the key generation algorithms, the main goal is to protect information sharing among these nodes and to simplify that process. To achieve this here, a cryptographic secret key generation algorithm is utilized. Unlike traditional networks with a central base station storing cryptographic keys, the decentralized system used in this study empowers all legitimate nodes to collaboratively generate and share the secret key. The secret key, essential to ensure communication channel confidentiality, is generated through pre-processing and post-processing functions, employing Gaussian prime numbers selected by each node \cite{authdata}.

\begin{figure}[] 
    \centering
    \includegraphics[width=0.45\textwidth]{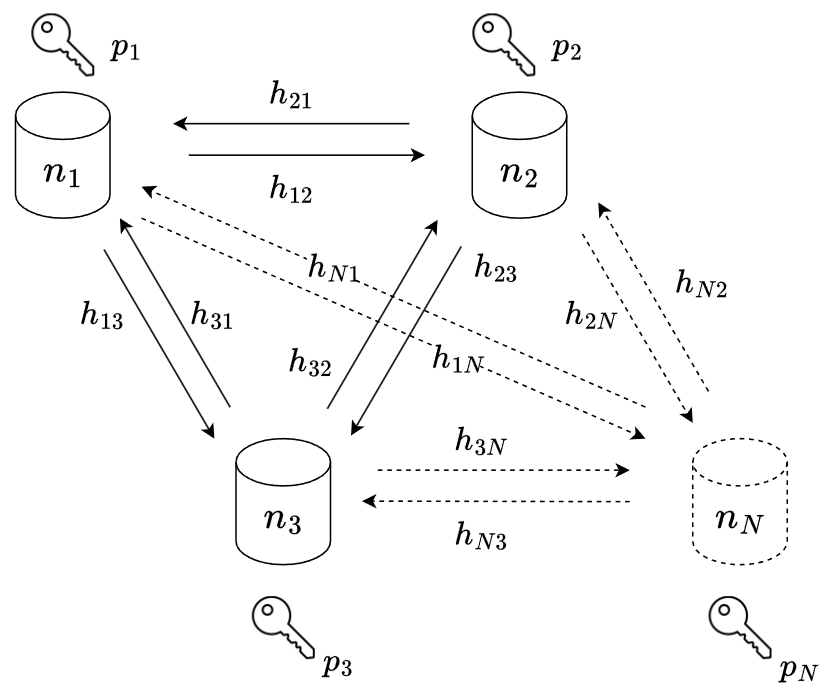}
    \caption{System Model}
    \vspace{-4mm}
    \label{system}
\end{figure}

As the details of the distributed system are investigated, as shown in Figure \ref{system}, the interaction of functions among nodes becomes clear. Each node, with its prime integer, computes and sends the processed output to all other nodes. The combined reception of these outputs establishes the groundwork for the next phase, resulting in a consistent secret key across the network \cite{scalablekey}.

The following sections explore processing functions, explaining how Analog Function Computation (AFC) separates individual input signals on the receivers. Afterward, the error model is examined, assessing how elements like thermal noise power, channel estimation errors, and the Rician factor affect the success of the key generation algorithm. This analysis results in a thorough channel model, covering large- and small-scale fading channels and accounting for reciprocal path loss.

Performance evaluation employs Monte Carlo simulations in a virtual environment, shedding light on the success rate for factorizing the secret key in the distributed wireless system. The simulation results comparing the success rates between 2-node and 3-node systems are presented in that section, along with an analysis of the impact of tolerance values defined in the factorization algorithms. Insights revealed in the discussion highlight path loss-induced large-scale fading as a dominant element impacting information and power loss.

This study not only reveals a conceptual framework but also substantiates its claims through rigorous mathematical formulations and simulations. The stage is set for advancing secure wireless systems, considering the interaction of processing functions, error models, channel characteristics, and performance evaluations.

\subsection{Literature Review}

Previous studies on AFC have garnered attention for its applications in various areas within wireless communication \cite{goldenbaum2012analog}, \cite{222222222}. This growing interest highlights the versatility and effectiveness of AFC in addressing various challenges in different contexts\cite{goldenbaum2013robust}. In \cite{altun2017testbed}, a centralized wireless network is designed and the pre-processing and post-processing functions measure the maximum and minimum performance of the channel there.
In \cite{authdata}, a secret key generation algorithm using AFC for the design of wireless sensor networks. In that study, the pre-processing and post-processing functions are utilized, and it is shown that the algorithm is resistant to spoofer attacks in certain scenarios. Following this study, in \cite{scalablekey}, a similar centralized wireless network is used to generate the secret key. In that study, the network is split to the layers a novel algorithm is announced in generating the secret key which is also resistant to the several attack schemes.
In the study \cite{huang2016channel}, a channel correlation based secret key generation experiment is conducted under eavesdropper for underwater acoustic systems. A comprehensive survey \cite{sahin2022survey} enlightens the capabilities and potentials of this technique with several use case examples.

Although several studies provide valuable insights into the applications and effectiveness of AFC in diverse wireless communication scenarios, there are notable gaps in the current literature that warrant further investigation. Investigating the feasibility and performance of AFC in real-world scenarios, including potential issues and practical constraints, would provide a more comprehensive understanding of its applicability. Also, for the algorithms using AFC, there is a need for more sophisticated and realistic error models that accurately capture the challenges posed by real-world communication channels.

The contributions of this study are the following
\begin{itemize}
    \item An error model for the secret key generation algorithm is demonstrated. The proposed distributed secret key generation algorithm is generalized for $N$ user case.
    \item The success rate of the multi-node network is obtained for ideal communication under different fading channels and thermal noise powers. Accordingly, the success rate is considered as upper bound performance for the corresponding case.
    \item The performance of the system is evaluated based on the success rate of the algorithm. The key components affecting the success rate are analyzed with certain values and intervals, and the optimal values to achieve a high success rate are explained while designing the system.
\end{itemize}

\section{System Model}The system consists of $N$ nodes and has a distributed network. Each node within a network cluster, comprising $N$ nodes, strives to establish $N-1$ secure mutual links with all other nodes. This security is ensured through the utilization of secret keys for node identification within a fading channel environment. The system model explains end-to-end key generation in transmission and key recovery in reception.

All nodes have full-duplex communication with each other, and the communication is carried out over the air. These nodes are symbolized in the set $N = \{n_1, n_2, \ldots, n_N\}$ where $n_i$ shows the $i$\textsuperscript{th} node. The locations of the nodes are fixed and stationary. Furthermore, the ideal system model includes stable thermal noise power for each node. \\

A secret key generation algorithm is designed for secure node-to-node transmission in a single network cluster. In this decentralized system, there is no base station that stores the secret key, and the secret key is generated and shared by all legitimate nodes within the network. To generate the secret key for the entire network, pre-processing and post-processing functions which are explained in the following section are utilized. Generating the secret key process begins with selecting a Gaussian prime number for each element inside the network. To provide this, the Gaussian prime number set is defined as $P = \{p_1, p_2, \ldots, p_N\} \subseteq C$. After a node selects its Gaussian prime integer, a pre-processing function utilizes it as a function of the prime integer. The specified node transmits the output value to every node within the network, excluding its own instance. After the transmission is completed, each node applies the post-processing function to generate the secret key \cite{scalablekey}. At the end of this process, each node has the same secret key value, which is the production of the complex Gaussian prime integers selected by the nodes. 

\subsection{Processing Functions}
Normally, in a system in which Analog Function Computation (AFC) is used, the input signals are superposed on the receivers. That means the individual input signals cannot be extracted from the output signal. However, by placing the pre-processing and post-processing functions in the transmitter and receiver, respectively, individual input signals can be identified in the receivers \cite{magicsurvey}. The pre-processing function in Equation (\ref{Eq1}) is determined as taking the natural logarithm of the input and then dividing the result by the estimated channel impulse response (CIR). In equation (\ref{Eq1}), the transmitter nodes are defined as $n_i$, and the CIR between the transmitter node and the receiver node is represented as $h_{ij}$.

\begin{equation}
\label{Eq1}
\begin{aligned}
&\color{black}{x} \in \mathbb{C}, \phi_i(x_i) = (\phi \circ x_i) \text{ of } n_i \\
&\phi_i(x_i) = \frac{1}{h_{ij}} \ln(x_i).
\end{aligned}
\end{equation}

The post-processing function (\ref{Eq2}) is determined as taking the exponential value of the input. In equation (\ref{Eq2}), the receiver nodes are defined as $n_j$.

\begin{equation}
\label{Eq2}
\begin{aligned}
&\color{black}y \in \mathbb{C}, \psi_j(y_j) = (\psi_j \circ y_j) \text{ of } n_j \\
&\psi_j(y_j) = e^{y_j}.
\end{aligned}
\end{equation}

\begin{figure}[] 
    \centering
    \includegraphics[width=0.5\textwidth]{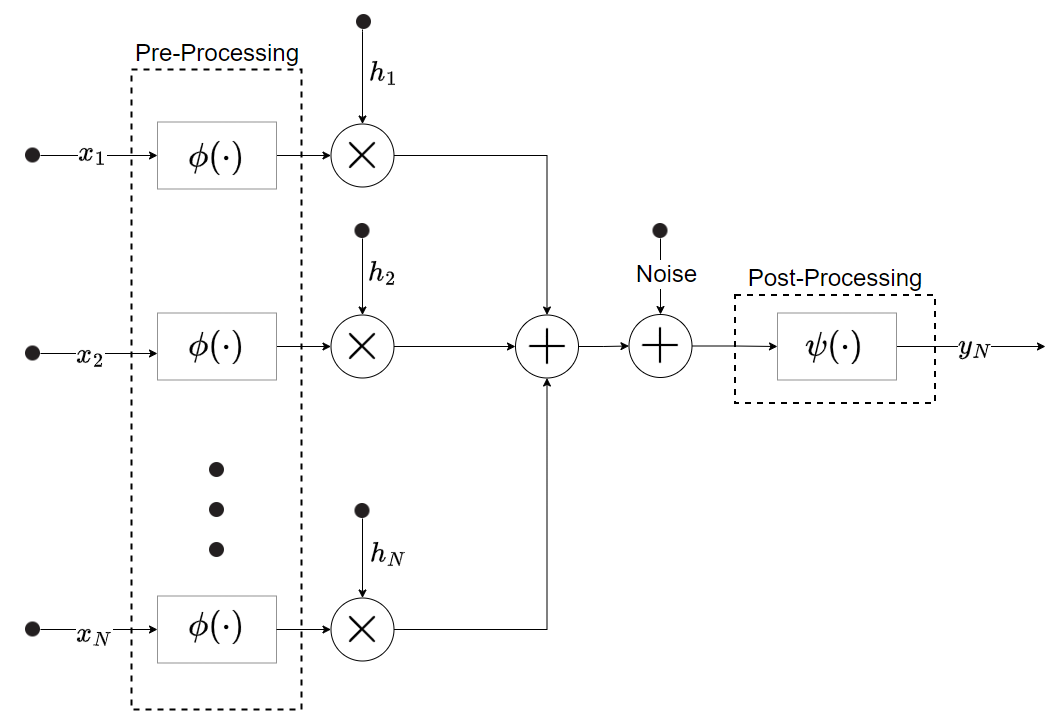}
    \caption{System Functions}
    \vspace{-4mm}
    \label{system-func}
\end{figure}

In the distributed system shown in Figure \ref{system-func}, each node calculates the output of the pre-processing function by using its own Gaussian prime integer and then sends it to all other nodes. Since there is a full-duplex communication between the nodes, all the nodes send and receive the output value of the pre-processing function at the same time. Thus, the CIR among the mutual nodes is the same when sending and receiving the data. After collecting the outputs of the pre-processing nodes, each node calculates the output of the post-processing function, which gives the secret key value \cite{authdata}. 

The secret key value acquired at each node can be decomposed into its constituent Gaussian prime integers. This process facilitates the determination of each node's unique key, thereby elucidating the association between messages and their respective nodes. Consequently, despite the use of superposition in AFC, this algorithm ensures the identification of all transmitter nodes along with their corresponding messages.

\section{Error Model}
The purpose of this study is to evaluate the errors of the proposed secret key generation algorithm for the distributed wireless network. There are several reasons that cause errors in generating the secret key in the nodes. The most effective ones are the thermal noise power in the medium, the difference between the estimated CIR and the ground truth CIR, and the Rician factor $K$. Based on these effective components, an error model is generated to evaluate the success of the key generation algorithm. 
The primary error model is formulated to assess the system's success rate. The success rate is calculated by identifying prime integers that correspond to the norms of Gaussian primes selected by the nodes. This is achieved through the factorization of the norm of the secret key value. Since the disturbances cause to get a different secret key value other than the theoretical secret key value, it is needed to decide a tolerance $\gamma_t$ in algorithms to try and find the prime integer values. The system first calculates the norms of both the theoretical secret key and the noisy secret key. It then breaks down the norm of the theoretical secret key value into its prime divisors. Using a defined tolerance level, it attempts to determine the prime divisors of the norm of the noisy secret key value and compares these findings with actual prime values.
The noisy key $S$ with $<5$ digits is factorized with trial divisions. For larger digits, Pollard's rho and $p-1$ methods are calculated numerically \cite{bach1991toward, pollard1974theorems}.
If the system successfully identifies the genuine prime divisors of the norm of the secret key within the specified tolerance range, it is considered a success, leading to an increase in the success rate.
\begin{algorithm}
    \SetKwInOut{Input}{Input}
    \SetKwInOut{Output}{Output}
    \underline{function Factorization} $(n)$\;
    \Input{A natural number $n$}
    \Output{A list of prime factors $a$}

    Set an empty $a$ array initializer\;
    $f \gets 2$ \Comment{Starting with the smallest prime factor}\;
    \While{$n > 1$}{
        \eIf{$n \mod f = 0$}{
            \Comment{Add $f$ to the list of prime factors}\;
            $n \gets n // f$ \Comment{Divide $n$ by $f$}\;
        }
        {
            $f \gets f + 1$ \Comment{Try the next potential prime factor}\;
        }
    }
    \Return{$a$}    
    \caption{Trial Division Algorithm}
    \label{algo:trial_division}
\end{algorithm}

In this test scenario, the noise in the medium is determined as a static value and the residual error on the estimated channel and $K$ of the channels are changed in certain intervals. For the estimated CIRs, the errors on these values are defined as $\tilde{h}^{ij} = h_e^{ij}/h^{ij},  \forall i \neq j$ where $h_e$ is the estimated CIR and $h$ is the ground truth CIR. 

It is assumed that the system follows pre-scheduled transmission for each node, leading to inter-user interference 0.  


\begin{algorithm}
    \SetKwInOut{Input}{Input}
    \SetKwInOut{Output}{Output}

    \underline{function Pre-Processing} $(p,h)$\;
    \Input{Two non-negative $p$ and $h$}
    \Output{$\frac{\log(p)}{h}$}
    
    \underline{function Post-Processing} $(y,\sigma)$\;
    \Input{Received signal $y$ and channel gain $h$}
    \Output{$\exp(y) + \sigma$}
    
    \underline{function Factorization} $(\psi)$\;
    \Input{$\psi \in \mathbb{R}$}
    \Output{factors}
    
    \underline{function NoisyFactorization} $(\psi, C)$\;
    \Input{$\psi \in \mathbb{R}$, $C \in \mathbb{Z}^+$}
    
        factors $\gets$ Factorization($\psi$)\\
        \eIf{$\left| \text{Re}(x)^2 + \text{Im}(x)^2 \right|$ is prime}
          {
            \textbf{return} factors\;
          }
          {
            \textbf{return} $\psi += \gamma_t$\;
          }
    
    \Output{Factors}

    \textbf{Initialize}: Gaussian Primes (GP), Array Size ($M$), $\sigma_n$\;
    \For{$i \gets 0$ \textbf{to} $2$} {
        \For{$j \gets 1$ \textbf{to} $T$} {
            Gaussian Primes $\gets$ GenerateRandomPrimes\;
            Success $\gets$ CalculateSuccessRate(GP, $\sigma_n$, $M$)\;
            StoreResults($i$, Success)\;
        }
    }
    \caption{Success Rate Algorithm}
    \label{algo:success_rate}
\end{algorithm}

During the simulations, the success rate is calculated for different channel estimation error values on the CIR for different $K$ values. The other aspects have minor effects on it, but they also need to be considered while designing the system.

In addition to the primary error model, an alternative error model is developed for the same system to evaluate the impacts of the distance between nodes. To calculate the success rate based on distance, the same input parameters are used, but the channel estimation error is given as a constant value which is $0.03$. The distance between the nodes is selected as the same for each interval. Then, the distance is changed during the simulations from the value of $1$ m to $150$ m. As given in the other error model, the success rate is measured for different $K$ and for each distance value. To determine the success rate, a tolerance value of $\gamma_t = 1500$ is specified for the factorization algorithms, and the system is modeled with two nodes and three nodes respectively.

\section{Channel Modeling}

Each user exposed to both large-scale and small-scale fading channel following Rice distribution as following 
\begin{equation}
f_i(x \mid \nu, \sigma)=\frac{x}{\sigma^2} \exp \left(\frac{-\left(x^2+\nu^2\right)}{2 \sigma^2}\right) I_0\left(\frac{x \nu}{\sigma^2}\right),
\end{equation}
where the $I_O$ is the zero order modified Bessel function of the first kind and the $K$ of the channel is defined as $K=\frac{\nu^2}{2 \sigma^2}$ with scale parameters $s>0$ and $\nu >0$. System follows homogeneous structure and channel is considered as reciprocal as $h_{ij} = h_{ji},\forall i\neq j$. The gain of the channel is assumed to be independent for each frame and is considered as constant within the frame. The channel coefficient with residual errors are modeled as $h_{ij}^e= h_{ij} + w_{h}$ where $w_h \sim \mathbb{N}(0,\sigma_h^2)$. As the path loss, each user's attenuation modeled with free space path loss equation as following \cite{balanis2016antenna}
\begin{equation}
\frac{P_r^j}{P_t^i}=G_t^i G_r^j\left(\frac{\lambda}{4 \pi d^{ij}}\right)^2
\end{equation}
where $P_t^i$ and $P_r^i$ are transmitted and received signal power of the $i-$th user, and $d^{ij}=|\rho_i - \rho_j|, \forall i\neq j$ where $\rho_{i}$ is the position vector of the $i-$th user, and $\lambda$ is the wavelength of the modulated carrier signal.  

Additionally, both Rayleigh and Rician fading were considered in the channels. Rayleigh fading occurs when no dominant line-of-sight component is present in the channel, resulting in the signal being subjected to multiple reflections and scattering from objects in the environment. This phenomenon is simulated by setting $K$ to 0, indicating the absence of a direct path between the transmitter and receiver. In order to unveil the impact of a stronger line-of-sight component, $K$ was gradually increased. By raising $K$ to 3 and subsequently to 20, the transition from a channel dominated by multipath scattering to one with a notable line-of-sight component was modeled. When $K=20$, the signal experienced a robust direct path between the transmitter and receiver, alongside scattered components \cite{rappaport}.

 \begin{figure}[] 
    \centering
    \includegraphics[width=0.45\textwidth]{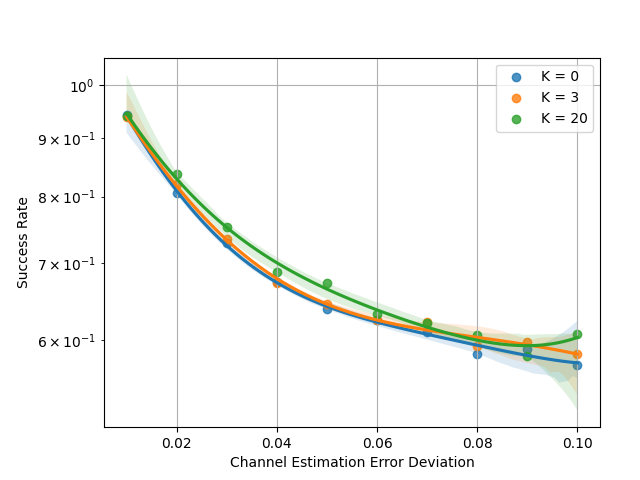}
    \caption {Success Rate versus Channel Estimation Error Deviation Analysis for a Two-Node.}
    \vspace{-4mm}
    \label{2-usr-err}
\end{figure}

 \begin{figure}[] 
    \centering
    \includegraphics[width=0.45\textwidth]{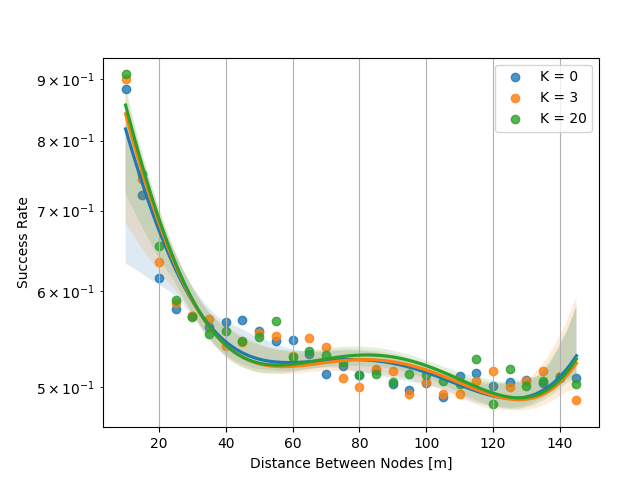}
    \caption{Success Rate versus Distance Between Nodes Analysis for a Two-Node.}
    \vspace{-4mm}
    \label{2-usr-dist}
\end{figure}

\section{Performance Evaluation}
The success rate for factorizing the norm of the secret key in the distributed wireless system is elaborately examined through extensive Monte Carlo simulations conducted in a virtual environment. Algorithm 2 outlines the complexity of the success rate calculation, where careful consideration is given to components such as the channel estimation error and $K$. The selected metrics, as highlighted in Table \ref{tab1}, play a crucial role in evaluating the system's robustness and reliability.

The simulation results, illustrated in Figure \ref{2-usr-err}, provide valuable information on the behavior of the system under varying conditions. In particular, the impact of the channel estimation error on the success rate is apparently observed, with an evident decrease as the error margin widens. This experimental observation indicates the system's sensitivity to accurate channel estimation for successful key factorization.

Additionally, an exciting result is observed in Figure \ref{2-usr-err} where the success rate reaches its highest values for $K=20$. This observation sheds light on the interaction between $K$ and the system's key factorization performance, suggesting an optimal range for achieving the highest success rate.

Insights from the alternative error model highlight the impact of inter-node distances on system behavior. Simulations show a decrease in the success rate with increasing node distance, as illustrated in Figure \ref{2-usr-dist}. It is observed that the success rate consistently stabilizes at $50\%$, regardless of $K$ values or increasing distance, indicating a minimum success floor value that all channels reach. This trend suggests a maximum feasible distance for effective key generation, prompting system designers to optimize parameters such as distance for reliable and secure wireless communication.

\begin{figure}[] 
    \centering
    \includegraphics[width=0.45\textwidth]{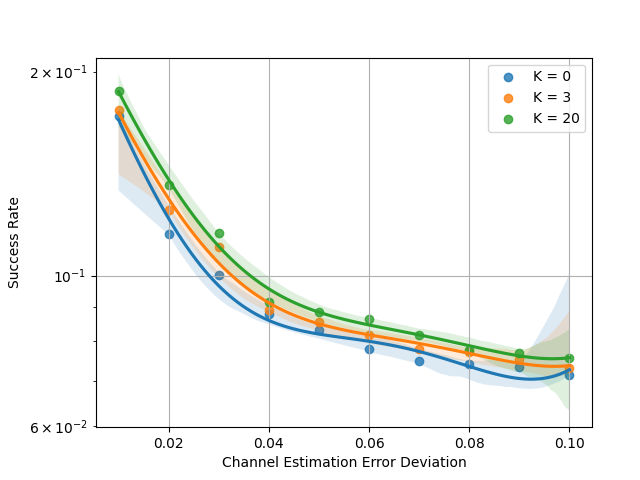}
    \caption {Success Rate versus Channel Estimation Error Deviation Analysis for Three-Node.}
    \vspace{-4mm}
    \label{3-usr-err}
\end{figure}

\begin{figure}[] 
    \centering
    \includegraphics[width=0.45\textwidth]{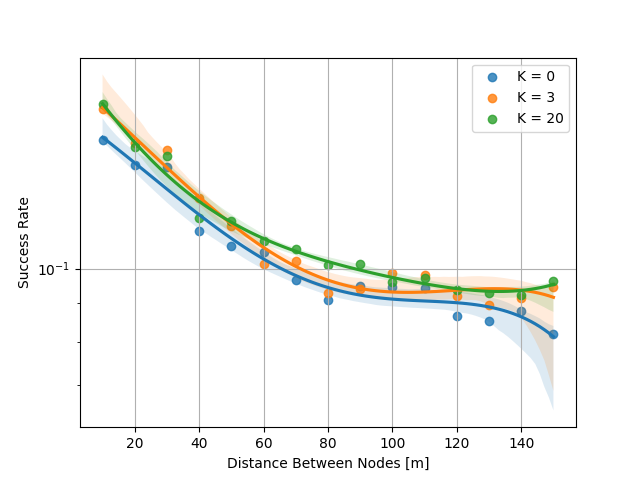}
    \caption{Success Rate versus Distance Between Nodes Analysis for Three-Node.}
    \vspace{-4mm}
    \label{3-usr-dist}
\end{figure}

After generating the two error models for systems with $2$ nodes, a new node was introduced to the systems to observe its impact on the success rate. The same tolerance value of $1500$ was used in the factorization algorithms to facilitate comparison. As a new node also selects a Gaussian prime integer as its key in the wireless system, the multiplication of prime integers increases, resulting in a larger secret key value. In addition to maintaining the same tolerance value in this new system, a dramatic decrease in the success rate was observed in both models. The simulation results in Figure \ref{3-usr-err} show that the maximum success rate is around $20\%$ for the smallest error estimation error and the maximum $K$. Similarly, the other result in Figure \ref{3-usr-dist} indicates that even when the nodes are brought closer together, a higher success rate is not achieved for the 3-node system. At the closest distance of $1$ meter, the success rate remains around $20\%$, with a continuous drop as the distance increases.

\begin{figure}[] 
    \centering
    \includegraphics[width=0.45\textwidth]{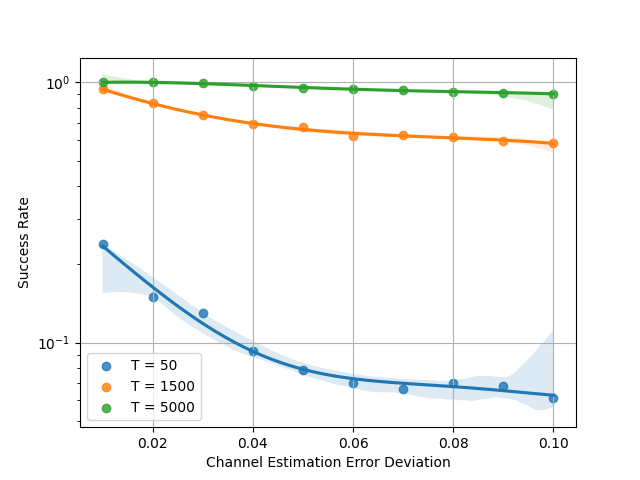}
    \caption {Success Rate versus Channel Estimation Error Deviation Analysis for a Two-Node: A Comparison of Tolerance Levels in Factorization Algorithms.}
    \vspace{-4mm}
    \label{tolerans-err}
\end{figure}

Moreover, the success rate is determined by applying various tolerance values in the factorization process of the secret key's norm. In Figure \ref{tolerans-err}, it has been observed that as the tolerance value increases, there is a significant rise in the success rate. This enhancement makes the system more robust against channel estimation errors and $K$, which exert a substantial influence on the success rate. However, this heightened tolerance also exposes the system to specific attack scenarios, as it increases the predictability of the secret key within a wider range for potential attackers. Consequently, selecting an optimal tolerance value becomes crucial to ensuring both the system's reliability and its resilience against potential attacks.

\begin{figure}[] 
    \centering
    \includegraphics[width=0.45\textwidth]{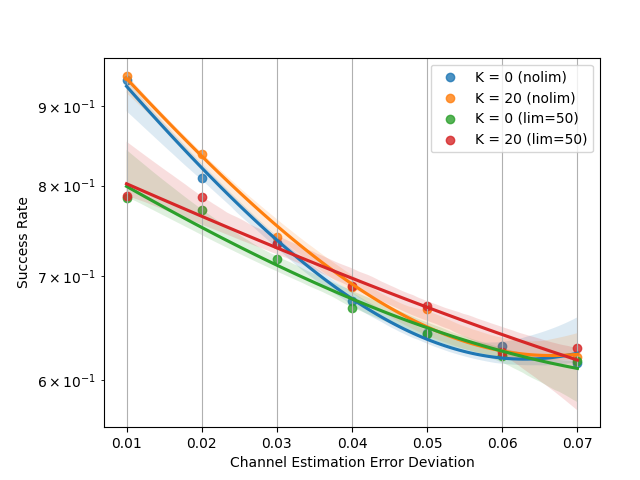}
    \caption {Success Rate versus Channel Estimation Error Deviation Analysis for a Two-Node: A Comparison of Unlimited Factorization Algorithm and Limited Factorization Algorithm.}
    \vspace{-4mm}
    \label{2-usr-lim}
\end{figure}

Furthermore, the system is subject to additional performance evaluation with a limited factorization algorithm. As depicted in Figure \ref{2-usr-lim}, when partial factorization (limited scenario) is applied under a specific channel estimation error, the success rate decreases. However, in areas where the estimation error is large, this disadvantage is mitigated, making it advantageous by removing the burden of full factorization and turning it into a favorable situation.
\balance

\subsection{Discussion}
It is observed that path loss caused by large-scale fading is the dominant reason of the information and power loss. Regarding other elements, such as channel estimation errors and thermal noise power, the system is resilient against thermal noise values. However, its performance is strictly bounded by channel estimation errors. Therefore, during the implementation of this algorithm and the design of the system, precise channel estimation becomes imperative. This requirement poses a significant challenge for researchers in this field.

\balance
\section{Conclusion}

In conclusion, a decentralized approach to secure wireless systems is revealed, emphasizing not only the collaborative efforts of the nodes inside the network but also the utilization of innovative algorithms for secret key generation. The distributed system, depicted in Figure \ref{system}, highlights the interaction of the pre-processing and post-processing functions, resulting in a secret key shared across the network. Examination of processing functions, particularly AFC, illustrates how individual input signals can be separated on the receivers by strategically placing pre-processing and post-processing functions at the transmitter and receiver, respectively. This insight lays the foundation for a robust and secure communication paradigm, where the shared secret key serves as the keystone for maintaining confidentiality.

The analysis of the error model reveals the nuanced impact of elements such as thermal noise power, channel estimation errors, and $K$ on the success of the key generation algorithm. A comprehensive channel model, integrating large scale and small scale fading channels while considering reciprocal path loss, enhances our understanding of the wireless environment. Through Monte Carlo simulations, the success rate in factorizing the secret key in the distributed wireless system is evaluated. The simulation results reveal that the 2-node system consistently demonstrates a higher success rate compared to the 3-node system. Additionally, it is observed that increasing the tolerance value in factorization algorithms leads to a significant increase in the success rate. The discussion underscores the prominence of path loss-induced large scale fading as a key contributor to information and power loss, shedding light on critical aspects for system designers.

This study bridges theoretical concepts with practical insights, substantiating claims through rigorous mathematical formulations and simulations. It not only presents a key generation system model, but also contributes to the advancement of secure wireless systems. Navigating the complex landscape of wireless communication, the decentralized framework, and innovative algorithms presented herein lay the foundation for future research and development in this domain. In our future study, we intend to extend the findings of this study by implementing them in real-world scenarios using Software Defined Radios (SDRs), thereby bridging the gap between theoretical analysis and practical applications.

 \begin{table}[]
   \centering
\caption{Input Parameters} 
  \resizebox{0.9\linewidth}{!}{
\label{tab1}
\begin{tabular}{|lclll|}
\hline
\multicolumn{5}{|c|}{\textbf{Parameters}}                                                                                                                \\ \hline \hline
\multicolumn{1}{|l|}{\textbf{Common Parameters}}      & \multicolumn{4}{c|}{\textbf{Values}}                                                                      \\ \hline
\multicolumn{1}{|l|}{Number of Simulations ($T$)}             & \multicolumn{4}{c|}{20000}                                                                  \\ \hline
\multicolumn{1}{|l|}{Factorization Tolerance ($\gamma_t$)}             & \multicolumn{4}{c|}{1500}                                                                  \\ \hline
\multicolumn{1}{|l|}{Number of Nodes ($N$)}                & \multicolumn{4}{c|}{2, 3}                                                                     \\ \hline
\multicolumn{1}{|l|}{Std. deviation of Thermal Noise Power ($\sigma_n$)}                   & \multicolumn{4}{c|}{0.01}                                                                        \\ \hline
\multicolumn{1}{|l|}{The Distance Between the Nodes ($d_{ij}$)}                   & \multicolumn{4}{c|}{15 m}                                                                        \\ \hline
\multicolumn{1}{|l|}{Channel Estimation Error ($\sigma_h$)}                   & \multicolumn{4}{c|}{0.01 to 0.1}                                                                        \\ \hline
\multicolumn{1}{|l|}{Rician Factor ($K$)}                     & \multicolumn{4}{c|}{0, 3, 20}                                                                          \\ \hline
\multicolumn{1}{|l|}{Carrier Frequency ($f_c$)}                   & \multicolumn{4}{c|}{2.4 GHz}                                                                        \\ \hline
\multicolumn{1}{|l|}{Channel Blocks of Length}                   & \multicolumn{4}{c|}{100000}                                                                        \\ \hline
\end{tabular}}
\vspace{-4mm}
\end{table}

\balance
\bibliographystyle{IEEEtran}

\end{document}